\documentclass[%
  reprint,
  superscriptaddress,
  amsmath,amssymb,
  aps,
  prd,
  nofootinbib
]{revtex4-2}

\usepackage{graphicx}
\usepackage{dcolumn}
\usepackage{bm}
\usepackage{braket}
\usepackage{booktabs}
\usepackage{array}
\usepackage{float}
\usepackage[colorlinks=true,linkcolor=blue,urlcolor=blue,citecolor=blue]{hyperref}
\usepackage{enumitem}
\usepackage{xcolor}

\begin{document}

\title{Quasiparticle Modeling of Strongly Magnetized Strange Quark Systems}

\author{Yogesh Kumar}
\affiliation{Department of Physics, Hansraj College, University of Delhi, Delhi - 110007, India}
\author{Manas Lohani}
\affiliation{Department of Physics, Hansraj College, University of Delhi, Delhi - 110007, India}
\author{Poonam Jain}
\email{pjain\_phy@aurobindo.du.ac.in (Corresponding Author)}
\affiliation{Department of Physics, Sri Aurobindo College, University of Delhi, Delhi - 110007, India}
\author{Pargin Bangotra}
\affiliation{Department of Physics, Netaji Subhas University of Technology, Dwarka, Delhi, India}
\author{Vinod Kumar}
\affiliation{Department of Physics, University of Lucknow, Lucknow, 226007, India}
\author{T. A. Nahool}
\affiliation{Physics Department, Faculty of Science, South Valley University, Qena, Egypt}
\author{Om Prakash}
\affiliation{Department of Physics, Sri Aurobindo College, University of Delhi, Delhi - 110007, India}
\author{Sanjay Tyagi}
\affiliation{Department of Physics, Sri Aurobindo College, University of Delhi, Delhi - 110007, India}

\date{\today}

\begin{abstract}
We examine spin-polarized Strange Quark Matter and the complex structure of Strange Quark Stars in the existence of a strong magnetic field. Our framework uses a quasiparticle model, enhanced by a QCD-informed effective quark mass dependent on both chemical potential value and magnetic field strength. The resulting Equation of State yields an energy per baryon below the iron limit, supporting the hypothesis that the actual ground state of matter is strange matter. The Tolman-Oppenheimer-Volkoff expressions are solved and find that the maximum stellar mass decreases with the degree of spin polarization, reaching as high as $2.35~M_\odot$ with a related radius $12.33$~km. Furthermore, the model's predictions for mass, radius, and surface redshift are in excellent conformance to observational limitations for the massive pulsar PSR J0952-0607. These results highlight the crucial role of quasiparticle medium effects and strong magnetic fields in describing dense matter.
\end{abstract}

\maketitle

\section{Introduction}
\label{sec:intro}

The study of compact stars offers a unique intersection of nuclear physics, particle physics, and astrophysics, which is used to test theories of matter under conditions of extreme density \cite{Baym:2018, Glendenning:2000}. Understanding the composition of these stellar remnants remains one of the major challenges in the area of quark matter. The theoretical framework for this search is Quantum Chromodynamics (QCD), whose phase diagram explores the different states of strongly interacting matter. The heavy-ion experiments probe the high-temperature regime at zero or low-density to study the complex structure of Quark-Gluon Plasma (QGP) \cite{Yogesh:2012, Yogesh:2015, Yogesh:2015a}, while the high-density and low-temperature region is the exclusive domain of compact stars \cite{Shapiro:1983}.
\\
Recent astronomical observations have offered essential limitations on the highly dense matter in case of equation of state (EoS). The discovery of massive pulsars, such as PSR J1614-2230 \cite{Fonseca:2016}, PSR J0348+0432 \cite{Antoniadis:2013}, and the more recent PSR J0740+6620 \cite{Cromartie:2020}, requires that the EoS be stiff enough to support stellar masses of at least two solar masses ($M_\odot$). In parallel, many compact objects are known to host extreme magnetic fields. Magnetars can exhibit surface fields of $10^{15}$ G, while their interiors may sustain fields several orders of magnitude higher, potentially reaching the $10^{18}$ G regime explored in this work. This favorable value of high mass and strong magnetic fields intensifies the search for a theoretical understanding of stellar interiors that can simultaneously account for both phenomena.
\\
An effective possibility, known as the Strange Matter Hypothesis \cite{Bodmer:1971, Witten:1984}, conjectures that matter composed of up, down, and strange quarks: Strange Quark Matter (SQM), may be the true ground state of matter. If correct, this implies the existence of Strange Quark Stars (SQSs), which would be self-bound by the strong interaction and exhibit macroscopic properties different from their traditional neutron star counterparts \cite{Alcock:1986, Farhi:1984}. This distinguishes SQSs from hybrid stars, which are gravitationally bound neutron stars that possess a quark matter core formed through a hadron-quark phase transition \cite{Sen:2021, Burgio:2002}.
\\
Phenomenological models such as the MIT bag model \cite{Chodos:1974je} have been extensively used to describe SQM. However, these models often depend on treating fundamental parameters like the quark masses as constants. A more physically relevant approach must account for in-medium QCD effects, where quark properties are expected to evolve with the energy scale. These effects are explained by quasiparticles. The concept of a density or chemical potential dependent "effective" mass has been explored in various contexts, from early models of confinement \cite{Fowler:1981} to modern quasiparticle \cite{wen:2009, Wen:2013} and density-dependent formalisms \cite{Chu:2014, Peng:1999, Bordbar:2012b}, and is an intrinsic feature of frameworks like the Nambu-Jona-Lasinio (NJL) model \cite{Buballa:2005}.
\\
Building on the spin-polarized SQM framework \cite{Bordbar:2011}, this work investigates the interplay of these effects by including a strong magnetic field. We introduce a key QCD modification that is an effective strange quark mass depends on both the in-medium chemical potential value and finite magnetic field strength. 
\newline 
In Section~\ref{sec:theory}, we detail this theoretical framework and derive the EoS. In Section~\ref{sec:structure}, we solve the famous Tolman-Oppenheimer-Volkoff (TOV) expressions to calculate the stellar mass-radius relations. Finally, in Section~\ref{sec:conclusion}, we outline our findings and demonstrate that our model's predictions are in remarkable agreement with the observed properties of the massive pulsar PSR J0952-0607, identifying it as a compelling SQS candidate.

\begin{figure*}[!ht]
    \centering
    \includegraphics[width=8.9cm, height=7cm]{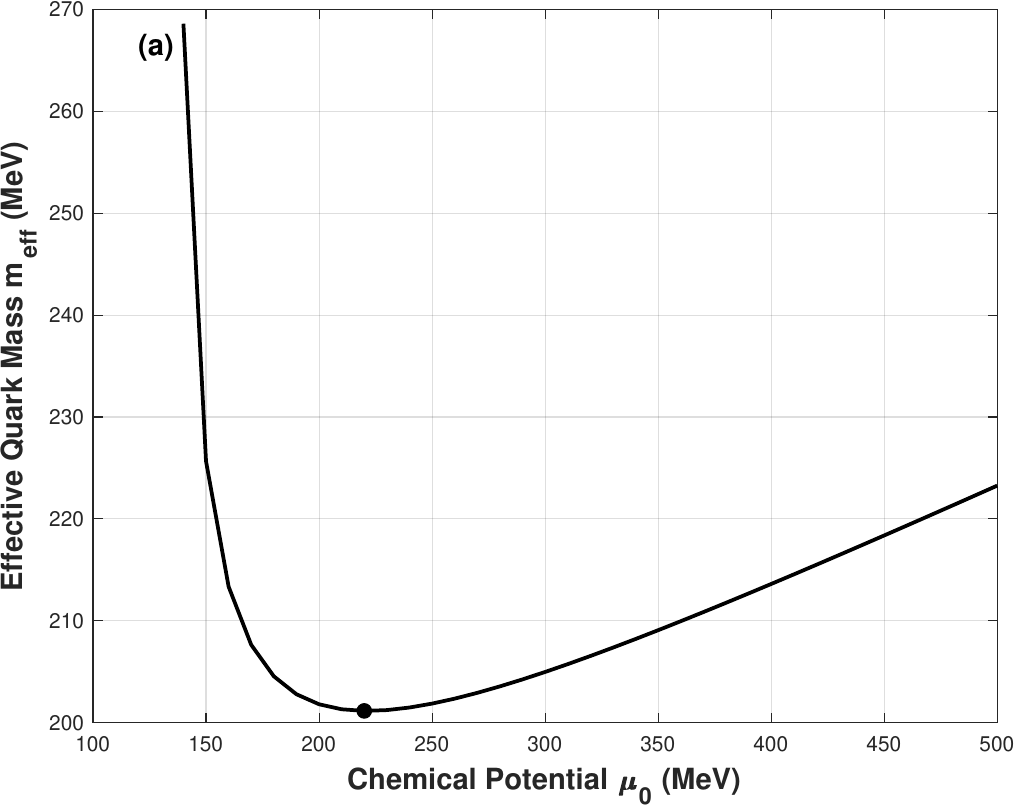}
    \includegraphics[width=8.9cm, height=7cm]{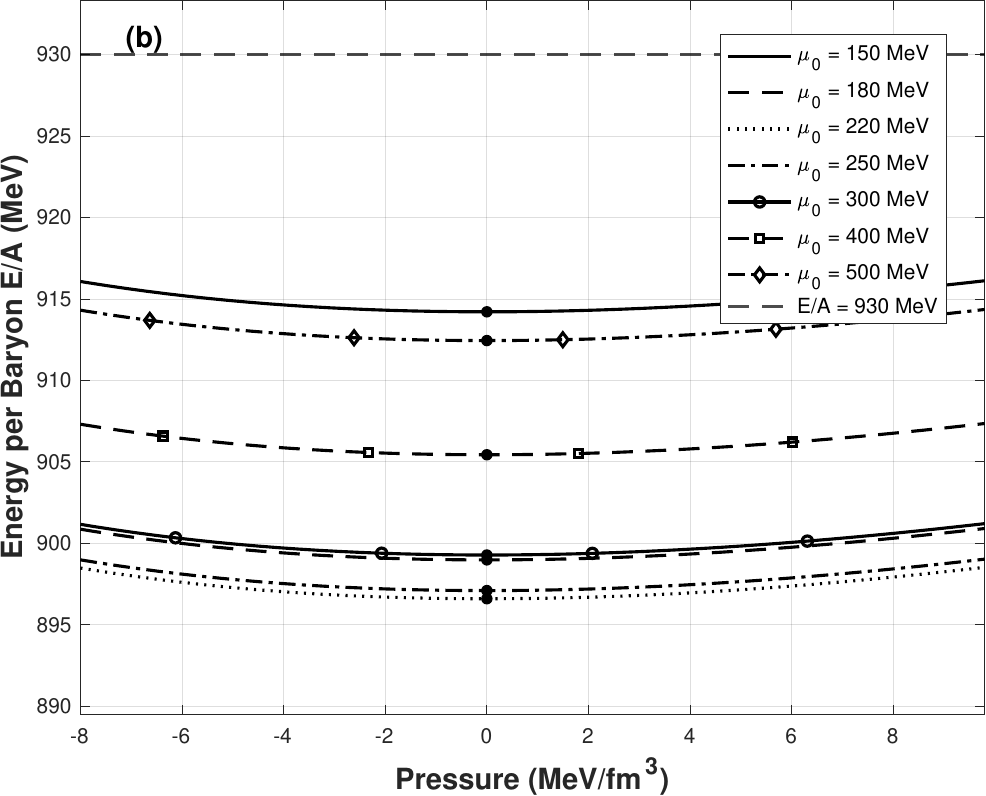}
    \caption{\textbf{Panel (a):} Variation of the effective quark mass ($m_{eff}$) is plotted with respect to the chemical potential ($\mu_0$). The black dot indicates the location of the minimum mass, corresponding to the most stable configuration. \textbf{Panel (b):} The energy per baryon ($E/A$) is plotted with pressure ($P$) for several chemical potential ($\mu_0$) value. The dashed horizontal line indicates the empirical nuclear saturation point at $E/A = 930$ MeV for comparison.}
    \label{fig:model_properties}
\end{figure*}

\section{Theoretical Analysis and Brief Model Discription of SQM}
\label{sec:theory}
The properties of strange quark matter (SQM) have been investigated by several authors using various methods \cite{Farhi:1984, Witten:1984, Alcock:1986, Alverdyan:2010, Li:2011}. The accelerators at RHIC and LHC provide us the signal of strangelets \cite{Weiner:2006}. In this work, we examine spin-polarized SQM composed of up quark, down quark, and strange quark. These particles are treated as fermions that possess an intrinsic quantum spin which can be oriented "up" ($+$) or "down" ($-$) relative to a quantization axis. The degree of spin alignment is quantified by the polarization parameter \cite{Bordbar:2011}, defined as:
\begin{equation*}
    \zeta_i = \frac{\rho_i^{(+)} - \rho_i^{(-)}}{\rho_i},
\end{equation*}
where $\rho_i = \rho_i^{(+)} + \rho_i^{(-)}$ is the total number density for a given quark flavor. While mathematically $\zeta$ can range from -1 to +1, the intrinsic is symmetric with respect to the choice of direction. By convention, we consider the net polarization to be positive, so our analysis is restricted to the range $0 \leq \zeta \leq 1$ without loss of generality. A value of $\zeta=0$ depicts an unpolarized state, while $\zeta=1$ defines a fully spin-aligned, ferromagnetic-like state. The primary astrophysical motivation for this study comes from the strong magnetic fields in magnetars, which are expected to induce such an alignment \cite{Chu:2014}. Therefore, this work directly incorporates a strong magnetic field to investigate the combined effects of magnetism and spin ordering on the properties of complex quark matter. Throughout this work, we assume the simplifying condition of equal polarization for all flavors, i.e., $\zeta_u = \zeta_d = \zeta_s \equiv \zeta$.

The present work follows the spin-polarized strange quark matter framework proposed by Bordbar and Peivand~\cite{Bordbar:2011}. Related studies by Bordbar and collaborators have subsequently employed the same effective treatment of spin polarization and magnetic interaction energy in magnetized neutron matter, nuclear matter, and strange quark matter at finite temperature~\cite{Bordbar:2011Neutron,Bordbar:2013SQM,Bordbar:2013Neutron,Bordbar:2014SQM,Bordbar:2016Nuclear,Rezaei:2017Nuclear,Bordbar:2022Neutron}.

It is important to note that there are two different theoretical approaches which are commonly employed in the literature for describing dense magnetized matter. One approach is based on the explicit treatment of orbital Landau quantization of charged particles, leading to discrete Landau levels and the associated anisotropic thermodynamic properties~\cite{Landau,Strickland,Felipe1,Felipe2,Sinha,Isayev,Ferraris2026,Chakrabarty1997,Lai1991,Broderick2000,Lu2022,Sinha2013}. Some of these reports follow the continuum approximation adopted as a reasonable approximation for highly dense quark matter. In such treatments, the Landau-level summation can approach a continuous momentum description under appropriate limiting conditions, providing a continuum approximation to the discrete spectrum~\cite{Chakrabarty1997,Lai1991,Broderick2000,Lu2022,Sinha2013}. 
\newline 
The second approach, adopted in the present work, is an effective thermodynamic framework in which the kinetic contribution is formulated within the continuum quasiparticle description, while the effects of the magnetic field are incorporated through spin polarization, magnetization, and the corresponding magnetic interaction energy~\cite{Bordbar:2011,Bordbar:2011Neutron,Bordbar:2013SQM,Bordbar:2013Neutron,Bordbar:2014SQM,Bordbar:2016Nuclear,Rezaei:2017Nuclear,Bordbar:2022Neutron}. Therefore, orbital Landau quantization is not introduced within the present formalism, since it belongs to a different microscopic description with a different set of physical assumptions. The objective of the present work is to investigate the influence of medium-dependent quasiparticle effects on the equation of state and the resulting macroscopic properties of magnetized strange quark matter within this effective thermodynamic framework.

Beyond the constant-mass approximation, our model introduces an effective quark mass dependent on the quark chemical potential ($\mu_0$) \cite{Schertler:1997, Pisarski:1989, Schertler:1997a}. This approach is physically motivated by in-medium QCD effects where quark properties are expected to evolve with the energy scale, a feature explored in various forms, including models with dynamical quark masses~\cite{Bordbar:2012}. While we neglect the masses of first two quarks; up ($u$) and down ($d$) as $m_u=m_d=0$, the in-medium effective quark mass squared ($m_{eff}^2$) is used from the quasiparticle model formalism. 

In the present formulation, the medium-dependent quasiparticle correction is introduced only for the strange quark sector. Since the objective is to investigate how the larger strange quark mass scale modifies the stability and structural properties of magnetized strange quark matter, the up and down quarks are retained within the conventional massless approximation commonly adopted in strange quark matter studies. This effective mass is strongly a function of chemical potential ~\cite{Srivastava:2010, LinLi:2021, Bannur:2008, Peshier:2000, Kumar:2018}:
\begin{equation}
\label{eq:ms_running}
    m_{eff}^2(\mu_0) = m_{s0}^2 + \sqrt{2} m_{s0} m_s( \mu_0) + m_s^2( \mu_0),
\end{equation}
where $m_{s0}$ is the current mass of the strange quark (150 MeV). The term $m_s$ represents the contribution of in-medium strange quark mass depends on the chemical potential \cite{wen:2009, Wen:2013}:
\begin{equation*}
    m_s(\mu_0) = \sqrt{\frac{g^2 \mu_0^2}{6\pi^2}}.
\end{equation*}
This approach is consistent with other quasiparticle descriptions of dense matter~\cite{Wen:2013, Peshier:2002}. The concept of a running coupling constant dependent on the chemical potential has been explored in various works~\cite{Wen:2010, Prasad:2001}. Considering the strong magnetic fields in highly dense matter, the running coupling constant $g$ is taken to be a function of both the chemical potential $\mu_0$ and the magnetic field $B$. For this work, the magnetic field is held constant at a value of $10^{18}$ G~\cite{Farias:2014, Wen:2025}:
\begin{equation}
\label{eq:g_coupling_mag}
    g^2(\mu_0,B) = \frac{48\pi^2}{(33 - 2N_f) \ln\left(0.8\frac{\mu_0^2}{\Lambda_1^2}\right) \left[1 + \alpha \ln\left(1 + \beta \frac{B}{\Lambda_2^2}\right)\right]}
\end{equation}
In this expression, $N_f$ is the number of active quark flavors. The scale parameters fixed as, $\Lambda_1 = 120$ MeV and $\Lambda_2 = 200$ MeV, with the free parameters adopted as $\alpha=2$ and $\beta=0.000327$~\cite{Farias:2014}. The logarithmic term in the denominator enforces the physical condition that its argument must be greater than one, leading to a lower bound on the chemical potential of $\mu_0 > \Lambda_1 / \sqrt{0.8}$.

The behavior of the effective quark mass ($m_{eff}$) is displayed as a function of the chemical potential in Panel (a) of Figure~\ref{fig:model_properties}. This curve reveals key physical insights into the nature of the dense quark medium. As low chemical potentials ($\mu_0$) approaching the QCD energy scale, the strong coupling constant, $g$, is large. This reflects a regime where non-perturbative QCD effects are dominant; the quarks are heavily "dressed" by their strong interactions with the area around plasma, resulting in a large effective mass.

As the chemical potential and density increase, we move into a regime of asymptotic freedom where the coupling constant decreases~\cite{Patra:1996}. This screens the interactions between quarks, causing their "dressing" to lessen and their effective mass to fall. This trend continues until a minimum is reached at $\mu_0 \approx 220$ MeV, marked by the black dot. This minimum represents the most energetically favorable (i.e., most stable) configuration for strange quark matter, a crucial condition for the validity of the Bodmer-Witten hypothesis.

Beyond this minimum, at very high chemical potentials, the $m_{eff}$ begins to rise again. Here, the system enters a perturbative QCD regime. Although the individual interactions are weaker (smaller $g$), the sheer density of the Fermi sea becomes the dominant factor. The rising $\mu_0^2$ term in Equation~(\ref{eq:ms_running}) represents the growing contribution of the dense medium to the quark's energy, causing its effective mass to increase.

\subsection{Equations of State}
\label{sec:eos_calc}
For all subsequent calculations, we adopt a bag constant, $B_{bag} = 55$ MeV/fm$^3$ and a constant magnetic field, $B = 10^{18}$ G. We employ a system of natural units where $\hbar = c = 1$, and all physical quantities are expressed in units of MeV. The presence of a strong magnetic field introduces a magnetic energy contribution, $E_M = - M \cdot B$, to the system~\cite{Bordbar:2011}. Assuming the magnetic field is uniform and oriented along the z-axis, this energy is given by:
\begin{equation}
    E_{M} = -\sum_{i=u,d,s} M_{z}^{(i)}B,
\end{equation}
where $M_{z}^{(i)}$ is the magnetization for each quark flavor $i$~\cite{Bordbar:2011}. The magnetization is defined as the total number of particles ($N_i$) multiplied by their magnetic moment ($\mu_i$) and the spin polarization parameter ($\zeta_i$):
\begin{equation}
    M_{z}^{(i)} = N_{i}\mu_{i}\zeta_{i}.
\end{equation}
From this, the magnetic energy contribution per unit volume, $\epsilon_M = E_M/V$, can be expressed in terms of the number density $\rho_i = N_i/V$~\cite{Bordbar:2011}:
\begin{equation}
\label{eq:e_mag}
    \epsilon_{M} = -\sum_{i=u,d,s} \rho_{i}\mu_{i}\zeta_{i}B.
\end{equation}
The total energy density ($\epsilon_{tot}^{(B)}$) of the magnetized system is the sum of the kinetic energy term, the bag constant, and this magnetic energy density~\cite{Bordbar:2011}:
\begin{equation}
\label{eq:etot_mag}
\begin{split}
    \epsilon_{tot}^{(B)} = & \frac{3}{16\pi^2} \sum_{p=\pm} \left[ k_F^{p} E_F^{p} (2(k_F^{p})^2 + m_{eff}^2) \right. \\
    & \qquad \left. - m_{eff}^4 \ln\left(\frac{k_F^{p} + E_F^{p}}{m_{eff}}\right) \right] + B_{bag} + \epsilon_{M} \\
    & + \frac{3\pi^{2/3}}{4} \rho^{4/3} \left[ (1+\zeta)^{4/3} + (1-\zeta)^{4/3} \right].
\end{split}
\end{equation}
The spin-dependent Fermi momentum, $k_F^{\pm}$, and Fermi energy, $E_F^{\pm}$, are given by:
\begin{align*}
    k_F^{\pm} &= (\pi^2\rho)^{1/3}(1 \pm \zeta)^{1/3}, \\
    E_F^{\pm} &= \sqrt{(k_F^{\pm})^2 + m_{eff}^2}.
\end{align*}
Consequently, the total pressure ($P$) includes a contribution from the magnetic field pressure $\frac{B^2}{8\pi}$~\cite{Bordbar:2011}:
\begin{equation}
    \label{eq:pressure_mag}
    P(\rho) = \rho \frac{\partial \epsilon_{tot}^{(B)}}{\partial \rho} - \epsilon_{tot}^{(B)} + \frac{B^2}{8\pi}.
\end{equation}

This pressure corresponds to the effective thermodynamic pressure associated with the present spin-polarized magnetized strange quark matter formalism following the framework of Bordbar and Peivand. The magnetic field contribution enters through the magnetic interaction energy density and magnetic pressure terms appearing in the equation of state. This set of equations now fully describes the magnetized strange quark matter equation of state used in our analysis.

Although the present work follows the formulation of Bordbar and Peivand, there are some useful studies reported that the rotational symmetry could broken by the magnetic field, leading to anisotropy between the longitudinal and transverse pressures~\cite{kohri,ferrer,martinez,chaichian}. Yet it is a matter of arguments where researchers claimed that the total pressure is truly isotropic so this issue is still a subject of ongoing debate~\cite{pote,ferr,dex}.  Therefore, by careful consideration of the current framework, the total pressure in this approach corresponds to the general thermodynamic pressure.

\begin{figure*}[!th]
    \centering
    \includegraphics[width=18cm, height=9cm]{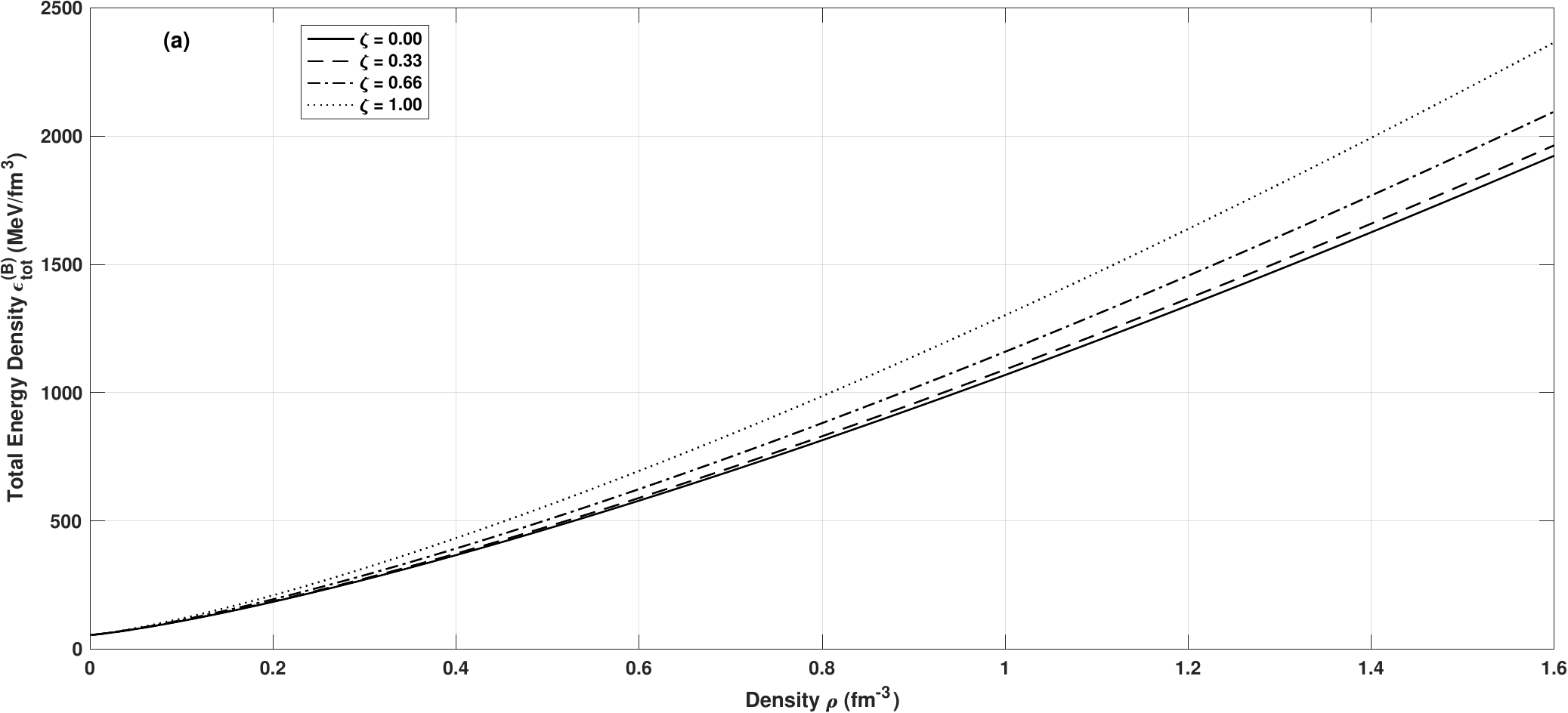}
    \includegraphics[width=18cm, height=9cm]{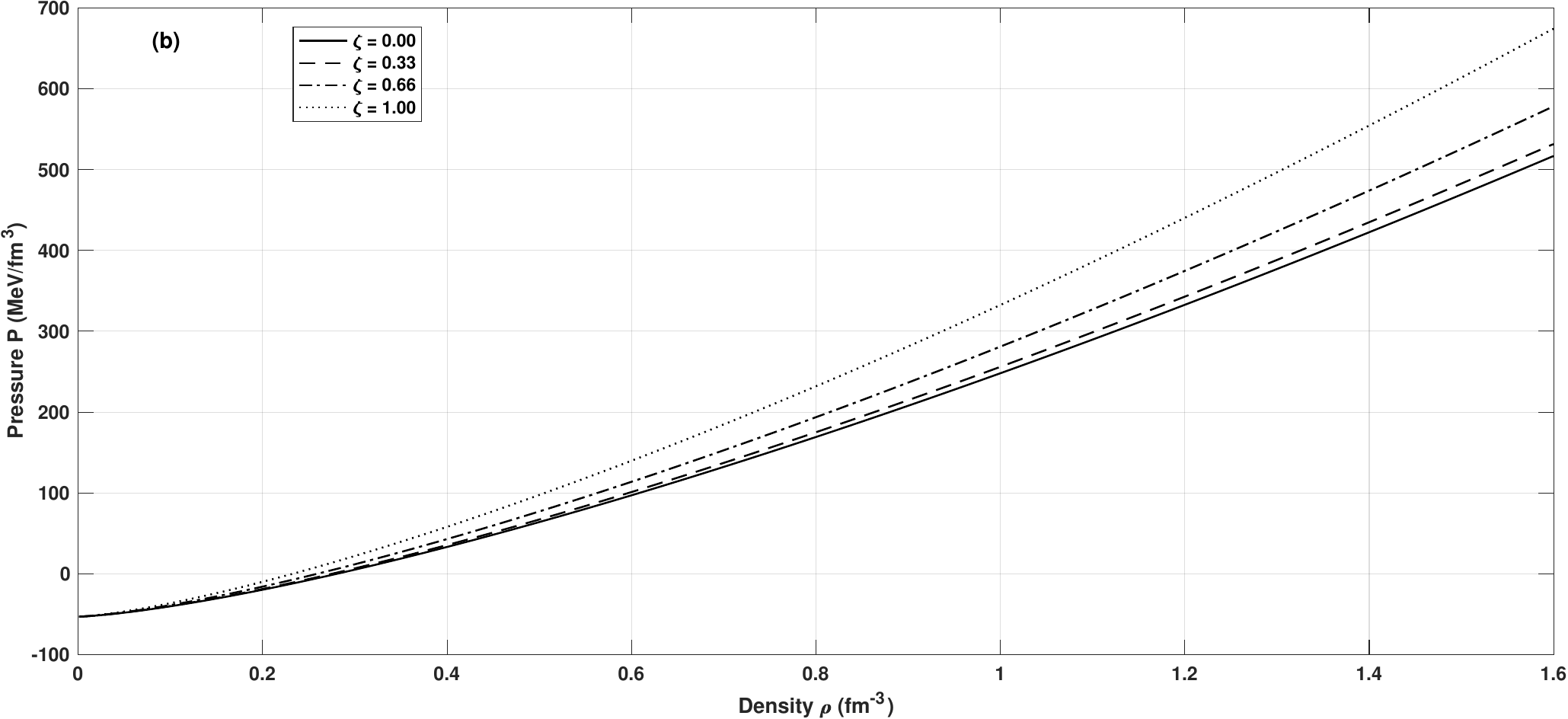}
    \caption{The equation of State for magnetized strange quark matter is shown. \textbf{Panel (a)} is shown the total energy density ($\epsilon_{tot}^{(B)}$) with baryon density ($\rho$). \textbf{Panel (b)} shows the pressure ($P$) versus baryon density ($\rho$) function. Each curve corresponds to a distinct spin polarization $\zeta$ values.}
    \label{fig:EoS_plots}
\end{figure*}

\subsection{Energy per Baryon and Stability}
\label{sec:stability}
To assess the stability of the system, we calculate the energy per baryon term, $E/A$, and the speed of sound squared, $c_s^2$. The energy per baryon is explained as the total energy density divided by the baryon number density:
\begin{equation}
\label{eq:E_per_A}
    E/A = \frac{\partial \epsilon_{tot}^{(B)}}{\partial \rho} - \frac{P}{\rho}.
\end{equation}
The most stable state of the system is found where $E/A$ reaches its lowest value. The speed of sound squared characterizes the stiffness of the EoS and is given by the fundamental relation \cite{Glendenning:2000}:
\begin{equation}
\label{eq:cs2}
    c_s^2 = \frac{\partial P}{\partial \epsilon_{tot}^{(B)}}.
\end{equation}
Here, $c_s^2 > 0$ ensures mechanical stability, while $c_s^2 \le 1$ is the relativistic causality constraint.

\begin{table}[t]
\centering
\large
\caption{Minimum energy per baryon term ($E/A$) is presented with chemical potential ($\mu_0$) value for magnetized strange quark matter.}
\label{tab:EoS_mag}
\begin{tabular}{c c}
\toprule
$\mu_0$ (MeV) & Min. $E/A$ (MeV) \\
\midrule
150 & 914.2167 \\
180 & 898.9884 \\
220 & 896.6036 \\
250 & 897.1046 \\
300 & 899.2883 \\
400 & 905.4419 \\
500 & 912.4513 \\
\bottomrule
\end{tabular}
\end{table}

\begin{table}[t!]
\centering
\large
\caption{Maximum value of the speed of sound squared value ($c_s^2$) is tabulated as a function of the spin polarization parameter ($\zeta$) for magnetized strange quark matter.}
\label{tab:SoS_mag}
\begin{tabular}{c c}
\toprule
$\zeta$ & Max. $c_s^2$ \\
\midrule
0.00 & 0.3168 \\
0.33 & 0.3176 \\
0.66 & 0.3196 \\
1.00 & 0.3231 \\
\bottomrule
\end{tabular}
\end{table}

\begin{figure*}[!th]
    \centering
    \includegraphics[width=8.9cm, height=7cm]{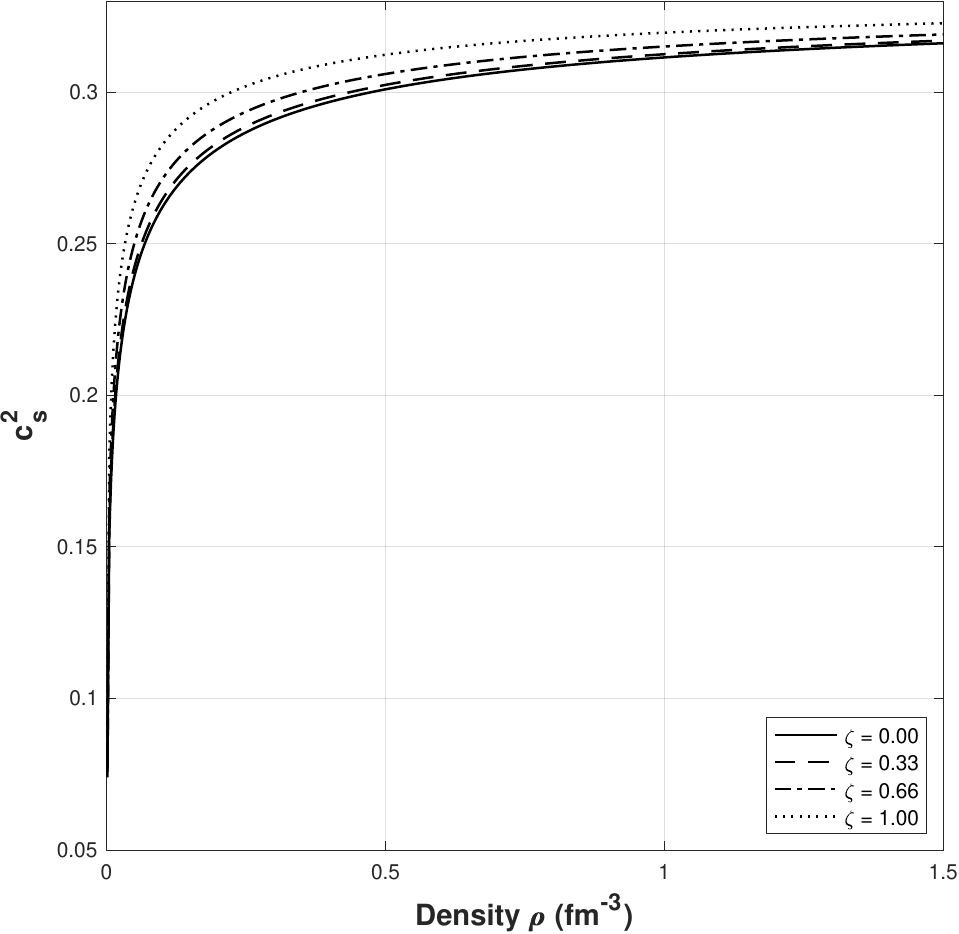}
    \caption{The speed of sound squared ($c_s^2$) value is plotted with baryon density ($\rho$) term for different spin polarization ($\zeta$) value.}
    \label{fig:cs2_vs_rho}
\end{figure*}

\begin{figure*}[!th]
    \centering
    \includegraphics[width=8.9cm, height=7cm]{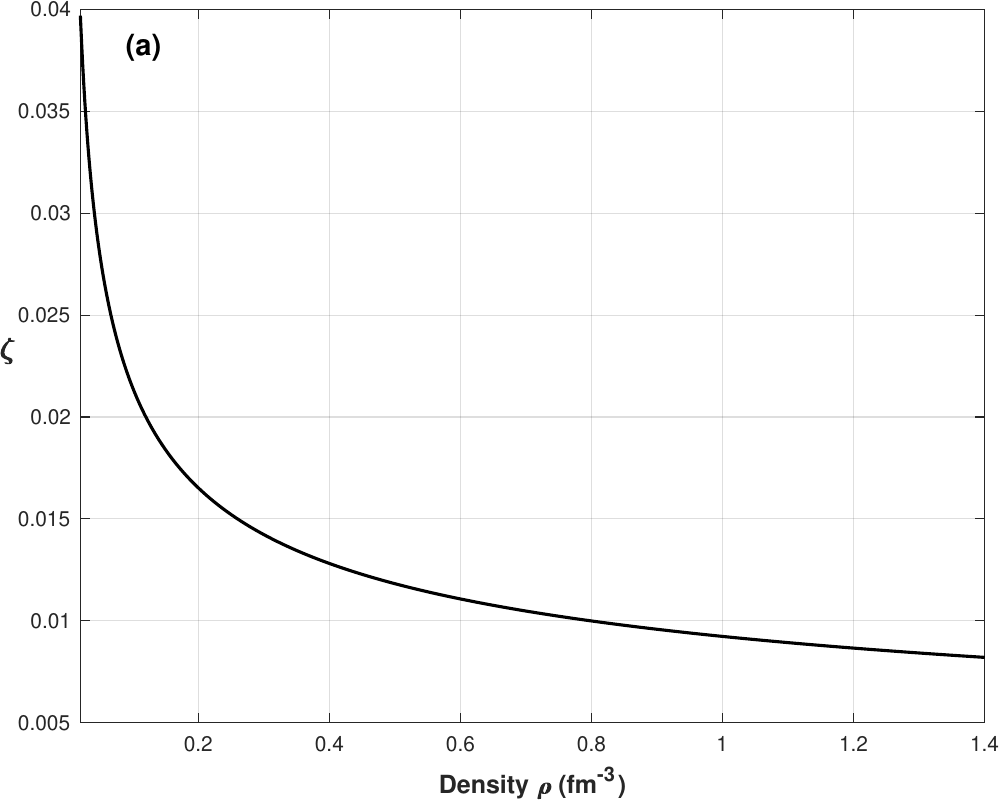}
    \includegraphics[width=8.9cm, height=7cm]{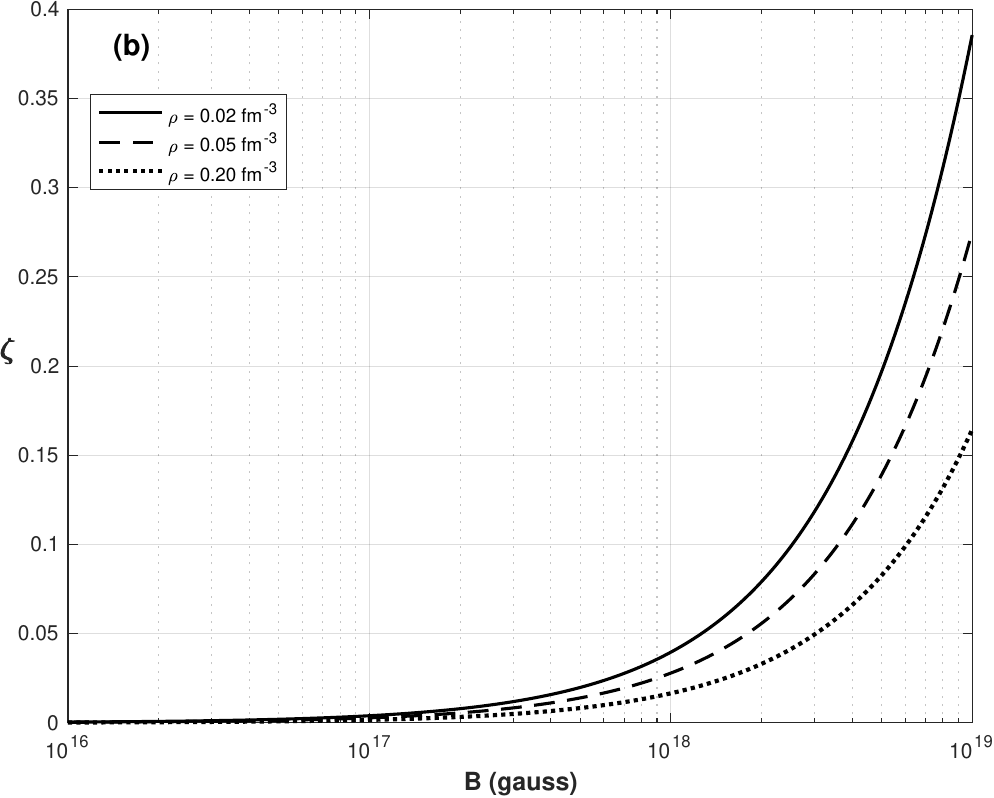}
    \caption{Behavior of the spin polarization parameter ($\zeta$) is displayed. \textbf{Panel (a)} shows $\zeta$ versus baryon density ($\rho$) function for a fixed magnetic field of $B=10^{18}$ G. \textbf{Panel (b)} shows $\zeta$ versus magnetic field strength ($B$) for several fixed densities.}
    \label{fig:zeta_behavior}
\end{figure*}

\begin{figure*}[!th]
    \centering
    \includegraphics[width=16cm, height=16cm]{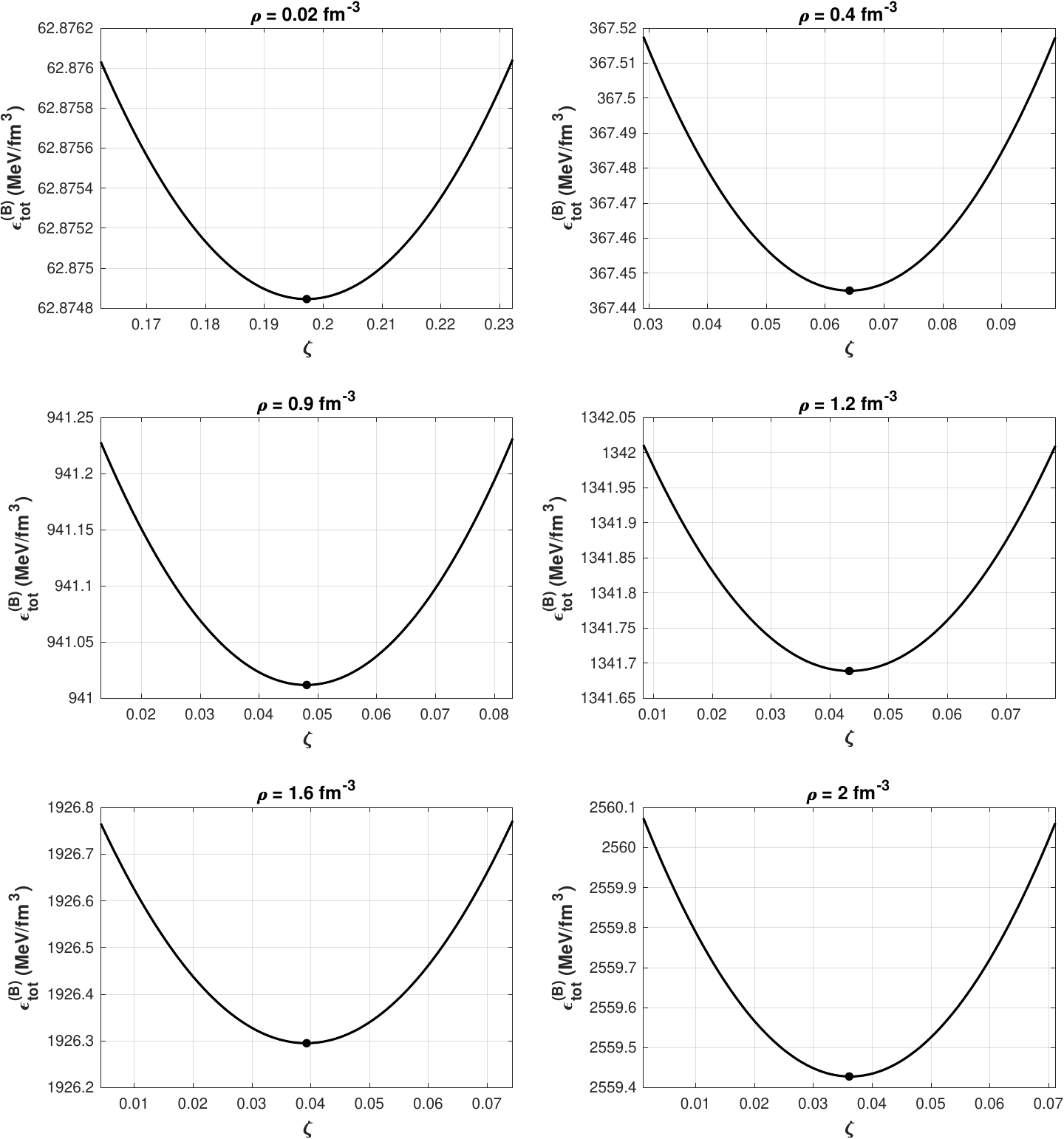}
    \caption{Total energy density ($\epsilon_{tot}^{(B)}$) is displayed with the spin polarization parameter ($\zeta$) for several fixed values of baryon density ($\rho$). The minimum of each curve, marked by a black dot.}
    \label{fig:E_vs_zeta}
\end{figure*}

In Panel (b) of Figure~\ref{fig:model_properties}, we plot the energy per baryon term with respect to pressure for various chemical potential values. The specific minimum values for each curve, which occur at zero pressure, are tabulated in Table~\ref{tab:EoS_mag}. By considering all configurations, the minimum energy per baryon lies significantly below that of iron (approximately 930 MeV, indicated by the dashed line). This result strongly supports the Bodmer-Witten hypothesis~\cite{Witten:1984}, indicating that the magnetized strange quark matter described by our model is absolutely stable and may be considered the true ground state of matter.
\\
The resulting EoS is shown in Figure~\ref{fig:EoS_plots}. Both the total energy density ($\epsilon_{tot}^{(B)}$) and the pressure ($P$) are shown to be monotonically increasing functions of the baryon density ($\rho$). This figure also clearly illustrates the effect of the spin polarization parameter, $\zeta$. As $\zeta$ increases from 0 (unpolarized) to 1 (fully polarized), both the energy density and pressure rise more steeply with baryonic density. This stiffening of the EoS is a direct result of the Pauli exclusion principle, as aligning more quark spins into a single state requires a greater energy cost. This has significant astrophysical implications, as a stiffer EoS provides greater resistance to gravitational collapse, directly affecting the structure of a SQS, as shown in Section~\ref{sec:structure}.
\\
The stiffness of the EoS is most precisely characterized by the speed of sound squared term ($c_s^2$), which is plotted as a function of density in Figure~\ref{fig:cs2_vs_rho}. The data confirm the trend observed in the EoS plots: for any given density, the value of $c_s^2$ increases as the spin polarization parameter $\zeta$ increases. The maximum values of $c_s^2$ for each polarization are listed in Table~\ref{tab:SoS_mag}. It is crucial to observe that all calculated values are well below the relativistic causality limit of $c_s^2 \le 1$ and approach the conformal limit of $c_s^2 \approx 1/3$ at high densities. These EoS results are used to model the stellar structure in the next section.
\\
To understand the behavior of the spin polarization, we analyze its dependence on density and magnetic field strength in Figure~\ref{fig:zeta_behavior}. Panel (a) shows that for our fixed magnetic field of $B=10^{18}$ G, the energetically preferred polarization, $\zeta$, is highest at low densities and decreases significantly as the matter becomes denser. This implies that while spin polarization may be relevant in the crust or the star's outer layers, the core matter will be nearly unpolarized. Panel (b) illustrates the expected causal relationship: a stronger magnetic field induces a greater degree of spin polarization at any given density.
\\
The two panels of Figure~\ref{fig:zeta_behavior} should be understood as illustrating two competing physical effects that determine the net spin polarization inside the star. Panel (b) shows the expected behavior: a stronger magnetic field provides a greater aligning force on the quark magnetic moments, leading to an increase in polarization. This effect would dominate for the constant density region. However, as one moves from a star's crust to its core, the baryon density increases dramatically. Panel (a) shows the outcome of this densification: Due to Pauli's exclusion principle, quarks with the same spin can't be packed together easily because it takes extra energy. To minimize the total kinetic energy, the system strongly prefers to equally fill the spin-up and spin-down states, which decreases the polarization. Thus, Panel (a) shows the net result of this competition, revealing that the effect of increasing density is dominant. Even in the existence of a strong background magnetic field, the Fermi pressure in the core drives the system towards an unpolarized state.
\\
Figure~\ref{fig:E_vs_zeta} explicitly demonstrates how the ground state of the system is determined. For each fixed baryon density, the total energy density, $\epsilon_{tot}^{(B)}$, exhibits a clear minimum at a particular spin polarization parameter value, $\zeta$. This minimum, marked by a black dot, represents the metastable state for that density. The observed plot confirms the trend seen previously in Ref. ~\cite{Bordbar:2011} where the position of this energy minimum shifts to lower values of $\zeta$ as the density increases, reinforcing that denser matter prefers a lower state of polarization. The exact values of these minima are presented in Table~\ref{tab:energy_minima}.

\begin{table}[h!]
\centering
\caption{Minimum total energy density ($\epsilon_{tot}^{(B)}$) and the corresponding spin polarization parameter ($\zeta$) for various baryon densities ($\rho$) in the existence of a strong magnetic field, $B=10^{18} G$ is presented in the table.}
\label{tab:energy_minima}
\begin{tabular}{c c c}
\toprule
$\rho$ (fm$^{-3}$) & Min. $\epsilon_{tot}^{(B)}$ (MeV/fm$^3$) & $\zeta$ \\
\midrule
0.02 & 62.8748  & 0.1972 \\
0.40 & 367.4450 & 0.0641 \\
0.90 & 941.0119 & 0.0481 \\
1.20 & 1341.6889& 0.0433 \\
1.60 & 1926.2953& 0.0393 \\
2.00 & 2559.4279& 0.0361 \\
\bottomrule
\end{tabular}
\end{table}

\section{Structure of Strange Quark Star}
\label{sec:structure}

The gravitational mass ($M$) term and radius ($R$) term are key observables of compact stars. We calculate these structural properties for our magnetized, spin-polarized SQS by numerically integrating the general relativistic equations of hydrostatic equilibrium-the Tolman-Oppenheimer-Volkoff (TOV) equations~\cite{Shapiro:1983}. In relativistic units ($c=1$) and retaining the gravitational constant $G$, the TOV equations are:
\begin{equation}
\label{eq:tov1}
    \frac{dm}{dr} = 4\pi r^2 \epsilon(r),
\end{equation}
\begin{equation}
\label{eq:tov2}
\begin{split}
    \frac{dP}{dr} = & -\frac{G m(r)\epsilon(r)}{r^2} \left( 1 + \frac{P(r)}{\epsilon(r)} \right) \\
    & \times \left( 1 + \frac{4\pi r^3 P(r)}{m(r)} \right) \left( 1 - \frac{2Gm(r)}{r} \right)^{-1},
\end{split}
\end{equation}
where $\epsilon(r)$ term and $P(r)$ term are the energy density factor and pressure term at a radius $r$. The term $m(r)$ is the gravitational mass enclosed within that radius:
\begin{equation*}
    m(r) = \int_0^r 4\pi r'^2 \epsilon(r') dr'.
\end{equation*}
The integration begins at the center of the star ($r=0$) with a selected central energy density, $\epsilon_c$, and the boundary condition $m(0)=0$. The integration proceeds outwards until the pressure vanishes, $P(R)=0$, which establishes star's radius $R$. The gravitational mass as a whole is then $M = m(R)$.
\\
Another important astrophysical observable is the surface gravitational redshift ($z_s$), which is determined by the star's compactness:
\begin{equation}
    \label{eq:redshift}
    z_s = \left(1 - \frac{2GM}{R c^2}\right)^{-1/2} - 1.
\end{equation}
We calculate this value using the final stellar parameters from our model in SI units.

\begin{figure*}[!th]
    \centering
    \includegraphics[width=16cm, height=6cm]{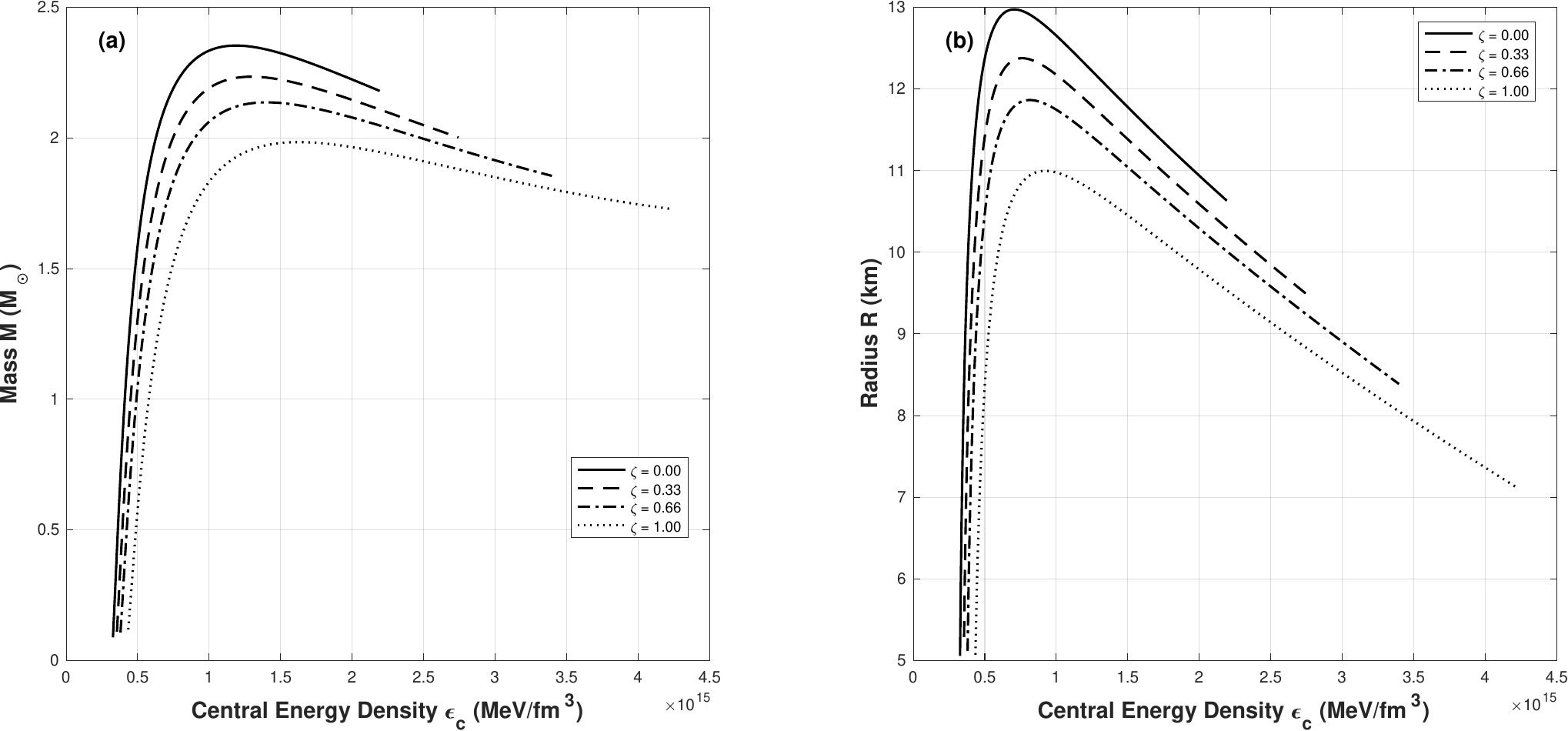}
    \caption{Stellar mass ($M$) in \textbf{panel (a)} and radius ($R$) in \textbf{panel (b)} is shown with central energy density ($\epsilon_c$). Each curve corresponds to a different spin polarization $\zeta$ value.}
    \label{fig:M_R_vs_Ec}
\end{figure*}

\begin{figure*}[!th]
    \centering
    \includegraphics[width=16cm, height=6cm]{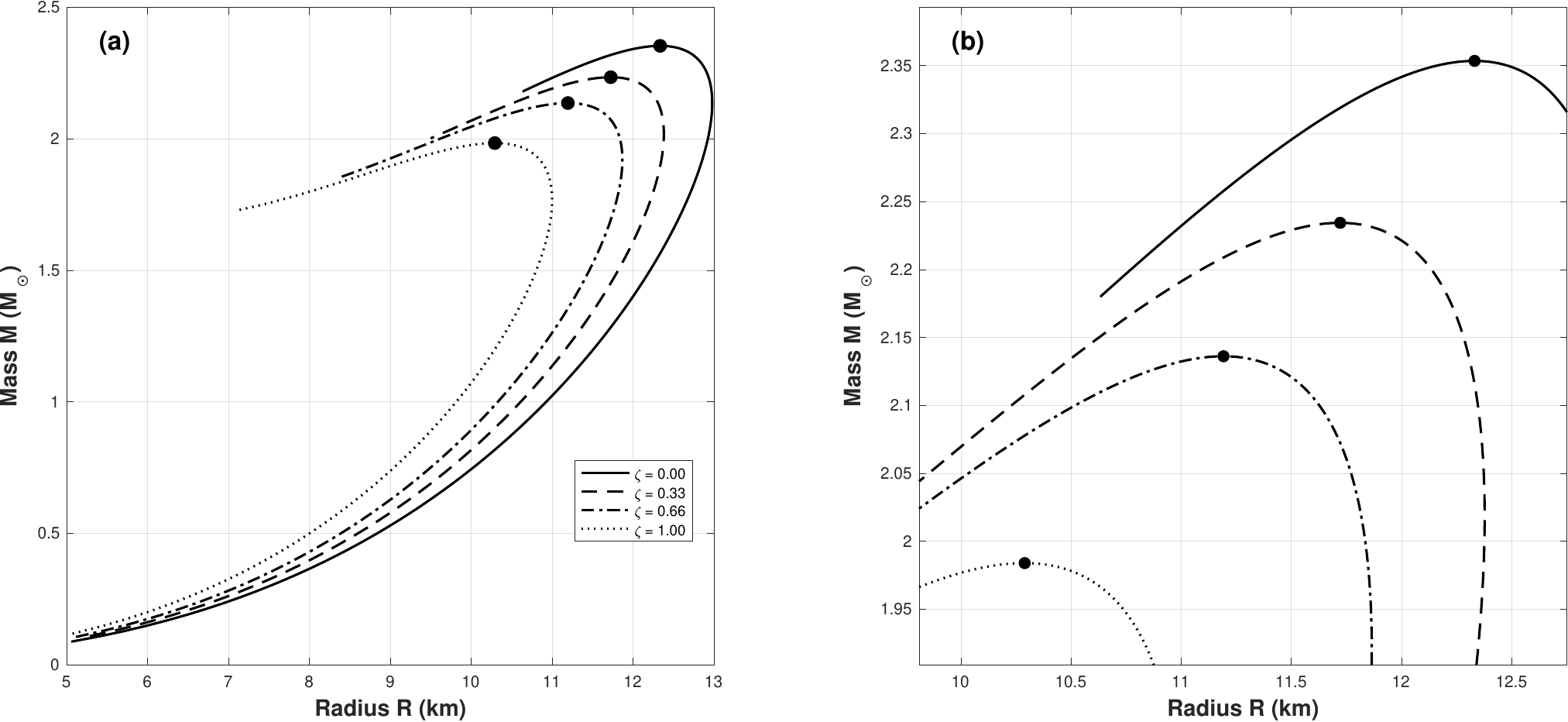}
    \caption{The mass-radius (M-R) relation for magnetized strange quark stars. Each curve represents a different spin polarization $\zeta$ value. The black dots indicate the maximum mass configuration for each sequence. \textbf{Panel (a)} shows the full stellar sequences, while \textbf{panel (b)} provides a zoomed-in view of the maximum mass region.}
    \label{fig:MR_relation}
\end{figure*}

\begin{table*}[t!]
\centering
\large
\caption{Maximum stellar mass ($M_{max}$), the corresponding radius ($R$), and surface redshift ($z_s$) is presented with the spin polarization parameter ($\zeta$) for magnetized strange quark stars.}
\label{tab:MRz_mag}
\begin{tabular}{c c c c}
\toprule
$\zeta$ & $M_{max}$ ($M_\odot$) & R (km) & $z_s$ (m/s) \\
\midrule
0.00 & 2.3536 & 12.3311 & 0.5139 \\
0.33 & 2.2344 & 11.7211 & 0.5127 \\
0.66 & 2.1362 & 11.1914 & 0.5140 \\
1.00 & 1.9840 & 10.2888 & 0.5241 \\
\bottomrule
\end{tabular}
\end{table*}

For each spin polarization $\zeta$ value, we perform the numerical integration of the TOV equations for a variety of central energy densities, $\epsilon_c$, to generate a corresponding family of unique star configurations. The resulting stellar mass and radius in relation to $\epsilon_c$ are shown in Figure~\ref{fig:M_R_vs_Ec}. These plots show the expected behavior. In this case, as the central density increases from low values, The star's radius and mass both increase until a maximum mass is reached. Beyond this peak, the configurations become unstable to gravitational collapse, and the radius begins to decrease. The final mass-radius (M-R) relations, which represent the sequences of stable stellar configurations, are displayed in Figure~\ref{fig:MR_relation}. The key macroscopic properties for the most massive stable star in each sequence are summarized in Table~\ref{tab:MRz_mag}.
\\
These plots encapsulate the main finding of our structural analysis. In contrast to models where polarization stiffens the EoS to support a larger mass, our model shows that an increase in the spin polarization parameter $\zeta$ results in a lower maximum stellar mass. This is evident as the M-R curve for $\zeta=1.00$ terminates at a significantly lower mass than the unpolarized ($\zeta=0$) case. This behavior, where spin polarization leads to a more compact star with a lower maximum mass, aligns with the findings of Bordbar and Peivand~\cite{Bordbar:2011} for magnetized quark stars.
\\
While this may seem counter-intuitive given that higher polarization leads to a locally stiffer Equation of State (as seen in Figure~\ref{fig:EoS_plots}), this result can be comprehended by taking into account the role of magnetic interaction energy (see Eqn.~\ref{eq:e_mag}). This term provides a negative contribution to the total energy density that grows with polarization. In the Tolman-Oppenheimer-Volkoff integration, this reduction in the total energy density for a given baryon density is the dominant effect. It effectively "softens" the EoS on a global, structural level, leading to a less massive and more compact stable star. Therefore, the increase in kinetic pressure from polarization is ultimately counteracted by the negative magnetic interaction energy, which governs the maximum mass of the star.
\\
Overall, our model predicts stable magnetized strange quark stars with maximum masses in the range of $1.98 - 2.35$ $M_\odot$, with corresponding radii between $10.29 - 12.33$ km and a calculated surface redshift between $0.5127 - 0.5241$ m/s. These predicted masses are notably high, capable of accounting for the most massive pulsars observed to date. For comparison, typical models for standard neutron stars constructed with purely hadronic equations of state often predict radii in the range of $12-13.5$ km \cite{Sen:2018}, while hybrid stars with a quark core can have maximum masses above $2.0 M_\odot$ with radii around $12$ km \cite{Sen:2021}. The properties predicted by our SQS model, particularly the ability to support a high mass within a moderately compact radius, allow for direct comparison with specific, high-mass astrophysical objects, as discussed in the final section.

\section{Summary and Conclusion}
\label{sec:conclusion}

In this work, we addressed various properties of strange quark matter (SQM) with spin-polarization and also the structure of strange quark stars (SQSs) in the presence of a high magnetic field. Our theoretical framework, detailed in Section~\ref{sec:theory}, is built upon a quasiparticle model that incorporates a running strange quark mass dependent on both factors; chemical potential and magnetic field strength, as described by Equations~(\ref{eq:ms_running}) and (\ref{eq:g_coupling_mag}). The equation of state (EoS) was subsequently defined in Subsection~\ref{sec:eos_calc}, where the total energy density term and pressure term were formulated to include magnetic contributions in Equations~(\ref{eq:etot_mag}) and (\ref{eq:pressure_mag}). Analysis of the energy per baryon, summarized in Table~\ref{tab:EoS_mag}, confirmed the stability of magnetized SQM. The EoS's stiffness, characterized by the speed of sound squared (Table~\ref{tab:SoS_mag}), was found to respect the relativistic causality limit. Finally, the Tolman-Oppenheimer-Volkoff (TOV) equations are solved to determine the macroscopic properties of the resulting SQSs, with the key findings for maximum mass, radius, and surface redshift presented in Table~\ref{tab:MRz_mag}.

To connect our theoretical predictions with astrophysical reality, we conducted a systematic survey of well-characterized massive pulsars. Our model predicts highly massive and compact objects with a mass between $1.98$ and $2.35$ $M_\odot$, a radius between $10.29$ and $12.33$ km, and a surface redshift between $0.5127$ and $0.5241$ m/s. This screening identified the pulsar PSR J0952–0607 as a prime observational counterpart~\cite{Romani:2022}. With an observed mass of $2.35 \pm 0.17$ $M_\odot$ and a radius near $12.25$ km, its properties align remarkably well with the upper limits of our model's predictions. The calculated surface redshift of $z_s \approx 0.517$ m/s for this object fits squarely within our predicted range, making it a compelling candidate for a magnetized strange quark star. 

A second plausible candidate, PSR J2215+5135, was also identified~\cite{Linares:2018}. While its mass of $2.27 ^{+0.17}_{-0.15}$ $M_\odot$ is an excellent fit, its measured radius is slightly larger than our model's upper bound. However, if its true radius lies at the lower end of its observational uncertainty, it also becomes a strong match. The ability of our model to account for such massive objects-a challenge for many standard nucleonic EoS models-underscores the importance of considering exotic phases of matter in stellar interiors. The extreme compactness demanded by these pulsars is naturally explained by the self-bound nature of strange quark matter, and our model provides a robust theoretical basis for this connection.
\newline 

We employ an effective quasiparticle thermodynamic framework for magnetized strange quark matter, following the formalism proposed by Bordbar and Peivand. Within this framework, the kinetic contribution is described using the continuum quasiparticle formulation, while the effects of the magnetic field are incorporated through spin polarization, magnetization, and the corresponding magnetic interaction energy. Consequently, the present model is formulated directly within this effective thermodynamic description rather than from an explicit orbital Landau-level formalism. The resulting equation of state provides a self-consistent description of stable magnetized strange quark matter and the corresponding strange quark stars within the scope of the adopted quasiparticle framework.
\newline 
The present work provides a basic investigation of spin-polarized strange quark matter within an effective quasiparticle framework. The resulting equation of state consistently predicts an energy per baryon below the iron stability limit, confirming the stability of magnetized strange quark matter considered in the present study. The outcomes also show an excellent consistency with observational constraints for the massive pulsar PSR J0952-0607 in terms of mass, radius, and surface redshift. Our results offer significant insights for both theoretical and experimental investigations of compact stars with quasiparticle medium effects.

Future work may extend the present effective thermodynamic framework by explicitly incorporating orbital Landau quantization together with the corresponding pressure anisotropy in the presence of strong magnetic fields. Such an extension would provide a more microscopic description of magnetized quark matter and enable a direct comparison between effective thermodynamic and fully Landau-quantized approaches. The present model may also be generalized to finite temperatures, allowing investigations of the thermodynamic behaviour of hot strange quark matter and the evolution of proto-quark stars within the quasiparticle framework. Furthermore, employing more sophisticated descriptions of quark matter, such as the Nambu-Jona-Lasinio (NJL) model or different versions of quasiparticle approach, could provide deeper insights into the phase structure of dense, magnetized QCD matter and further refine the equation of state.

\section*{Acknowledgements}

The Department of Physics at Hansraj College, University of Delhi, is acknowledged by the authors for providing the favorable research environment needed to complete this study. 
The authors would also like to express their gratitude to Prof. Abdel Nasser Tawfik for his constructive suggestions in refining and improving the manuscript.

\end{document}